\documentclass[12pt]{article}
\usepackage{epsfig}

\usepackage{slashed}
\newcommand{\notp}{{\slashed{p}}}

\newcommand{\re}{\mathop{\mathrm{Re}}\nolimits}
\newcommand{\im}{\mathop{\mathrm{Im}}\nolimits}

\begin{document}

\title{
\vskip-3cm{\baselineskip14pt
\centerline{\normalsize DESY 08--157\hfill ISSN 0418-9833}
\centerline{\normalsize NYU--TH/08/12/31\hfill}
\centerline{\normalsize December 2008\hfill}}
\vskip1.5cm
\bf A Novel Formulation of Cabibbo-Kobayashi-Maskawa Matrix Renormalization}

\author{Bernd A. Kniehl$^*$ and Alberto Sirlin$^\dagger$\\
\\
{\normalsize\it $^*$ II. Institut f\"ur Theoretische Physik, Universit\"at
Hamburg,}\\
{\normalsize\it Luruper Chaussee 149, 22761 Hamburg, Germany}\\
\\
{\normalsize\it $^\dagger$ Department of Physics, New York University,}\\
{\normalsize\it 4 Washington Place, New York, New York 10003, USA}}

\date{}

\maketitle

\begin{abstract}
We present a gauge-independent quark mass counterterm for the on-shell
renormalization of the Cabibbo-Kobayashi-Maskawa (CKM) matrix in the Standard
Model that is directly expressed in terms of the Lorentz-invariant self-energy
functions, and automatically satisfies the hermiticity constraints of the mass
matrix.
It is very convenient for practical applications and leads to a
gauge-independent CKM counterterm matrix that preserves unitarity and
satisfies other highly desirable theoretical properties, such as flavor
democracy.

\medskip

\end{abstract}

\newpage

\begin{figure}[ht]
\begin{center}
\includegraphics[bb=112 626 524 779,width=\textwidth]{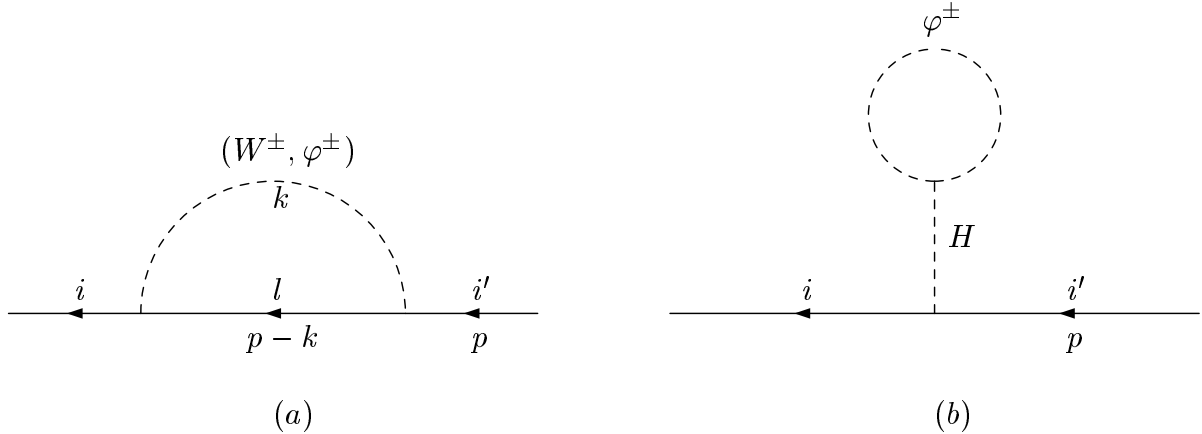}
\end{center}
\caption{\label{fig:one}%
Fermion self-energy diagrams.}
\end{figure}

Recently, a new approach to renormalize the Cabibbo-Kobayashi-Maskawa (CKM)
matrix \cite{Cabibbo:1963yz} at the one-loop level in the Standard Model (SM)
framework was proposed \cite{Kniehl:2006bs,Kniehl:2006rc}.
It is based on a simple procedure to separate the external-leg mixing
corrections generated by the Feynman diagrams of Fig.~\ref{fig:one} into
gauge-independent {\it self-mass} (sm) and gauge-dependent
{\it wave-function renormalization} (wfr) contributions, and to
adjust non-diagonal counterterm matrices to cancel the sm contributions,
subject to constraints imposed by the hermiticity of the mass matrices.
Diagonalization of the complete mass matrices for up-type and down-type quarks
leads then to an explicitly gauge-independent CKM counterterm matrix that
preserves unitarity.

In this paper we discuss an alternative on-shell approach that presents
especially attractive features.
It is a based on a gauge-independent mass counterterm matrix that is directly
expressed in terms of the Lorentz-invariant self-energy functions and
automatically satisfies the hermiticity constraints of the mass matrix.

On covariance grounds, the self-energy $\Sigma_{ii^\prime}(\notp)$ associated
with Fig.~\ref{fig:one} is of the form
\begin{equation}
\Sigma_{ii^\prime}(\notp)=\notp a_-\Sigma_{ii^\prime}^L(p^2)
+\notp a_+\Sigma_{ii^\prime}^R(p^2)
+a_-A_{ii^\prime}^L(p^2)+a_+A_{ii^\prime}^R(p^2),
\label{eq:sig}
\end{equation}
where $a_\pm=(1\pm\gamma_5)/2$ are the chiral projectors and
$\Sigma_{ii^\prime}^{L,R}(p^2)$ and $A_{ii^\prime}^{L,R}(p^2)$ are the
invariant self-energy functions.
At one loop in the SM, we have
\begin{equation}
m_{i^\prime}A_{ii^\prime}^L(p^2)=m_iA_{ii^\prime}^R(p^2).
\label{eq:LR}
\end{equation}
Explicit one-loop expressions for the SM in the $R_\xi$ gauges are given in
the Appendix of Ref.~\cite{Kniehl:2000rb} in combination with the tadpole
contributions in Eq.~(B.3) of Ref.~\cite{Kniehl:1990mq} and Eq.~(A5) of
Ref.~\cite{Hempfling:1994ar}.

The corresponding self-energy corrections to an external leg involving an
outgoing quark is
\begin{equation}
\Delta\mathcal{M}_{ii^\prime}^\mathrm{leg}
=\overline{u}_i(p)\left[\Sigma_{ii^\prime}(\notp)-\delta m_{ii^\prime}\right]
\frac{1}{\notp-m_{i^\prime}},
\label{eq:DelM}
\end{equation}
where $i$ denotes the flavor of the external quark of mass $m_i$ and
four-momentum $p$, $i^\prime$ that of the virtual quark of mass $m_{i^\prime}$,
and $\delta m_{ii^\prime}$ is the mass counterterm matrix.
For definiteness, we first consider the case in which $i$ and $i^\prime$ in
Fig.~\ref{fig:one} are up-type quarks and $l$ in the loop is a down-type
quark.
In this case, the proposed mass counterterm is
\begin{eqnarray}
\delta m_{ii^\prime}&=&V_{il}V_{li^\prime}^\dagger
\re\left\{a_+\left[\frac{m_{i^\prime}}{2}\tilde{\Sigma}_{ii^\prime}^L(m_i^2)
+\frac{m_i}{2}\tilde{\Sigma}_{ii^\prime}^R(m_i^2)
+\tilde{A}_{ii^\prime}^R(m_i^2)\right]\right.
\nonumber\\
&&{}+\left.a_-\left[\frac{m_i}{2}\tilde{\Sigma}_{ii^\prime}^L(m_{i^\prime}^2)
+\frac{m_{i^\prime}}{2}\tilde{\Sigma}_{ii^\prime}^R(m_{i^\prime}^2)
+\tilde{A}_{ii^\prime}^L(m_{i^\prime}^2)\right]\right\},
\label{eq:delm}
\end{eqnarray}
where $\tilde{\Sigma}_{ii^\prime}^{L,R}(p^2)$ and
$\tilde{A}_{ii^\prime}^{L,R}(p^2)$  are the invariant self-energies with
$V_{il}V_{li^\prime}^\dagger$ factored out and, following standard
conventions, $V_{il}$ is the CKM matrix element involving the up-type quark
$i$ and the down-type quark $l$.

An explicit expression for $\delta m_{ii^\prime}$ in the SM can be obtained by
using Eq.~(21) of Ref.~\cite{Kniehl:2006rc}, which provides the Feynman
amplitude $M_{ii^\prime}^{(1)}$ corresponding to Fig.~\ref{fig:one} in the
$R_\xi$ gauges.
Recalling that $\Sigma_{ii^\prime}(\notp)=iM_{ii^\prime}^{(1)}$ and taking
into account Eq.~(\ref{eq:sig}), one can readily determine the contributions
of each term of Eq.~(21) to the invariant functions and, via
Eq.~(\ref{eq:delm}), to $\delta m_{ii^\prime}$.
One finds that only the first three terms of Eq.~(21) give non-vanishing
contributions to $\delta m_{ii^\prime}$.
Separating out the chiral components,
\begin{equation}
\delta m_{ii^\prime}=a_+\delta m_{ii^\prime}^{(+)}
+a_-\delta m_{ii^\prime}^{(-)},
\label{eq:delmpm}
\end{equation}
we obtain the SM expressions
\begin{eqnarray}
\delta m_{ii^\prime}^{(+)}&=&\frac{g^2}{32\pi^2}V_{il}V_{li^\prime}^\dagger
\re\left\{m_i\left(1+\frac{m_i^2}{2m_W^2}\Delta\right)
-\frac{m_{i^\prime}m_l^2}{2m_W^2}[3\Delta+I(m_i^2,m_l)+J(m_i^2,m_l)]
\right.
\nonumber\\
&&{}+\left.m_{i^\prime}\left(1+\frac{m_i^2}{2m_W^2}\right)
[I(m_i^2,m_l)-J(m_i^2,m_l)]
\right\},
\label{eq:delmp}\\
\delta m_{ii^\prime}^{(-)}&=&\frac{g^2}{32\pi^2}V_{il}V_{li^\prime}^\dagger
\re\left\{m_{i^\prime}\left(1+\frac{m_{i^\prime}^2}{2m_W^2}\Delta\right)
-\frac{m_im_l^2}{2m_W^2}[3\Delta+I(m_{i^\prime}^2,m_l)+J(m_{i^\prime}^2,m_l)]
\right.
\nonumber\\
&&{}+\left.m_i\left(1+\frac{m_{i^\prime}^2}{2m_W^2}\right)
[I(m_{i^\prime}^2,m_l)
-J(m_{i^\prime}^2,m_l)]\right\},
\label{eq:delmm}
\end{eqnarray}
where $g$ is the SU(2) gauge coupling, 
$\Delta=1/(n-4)+[\gamma_E-\ln(4\pi)]/2+\ln(m_W/\mu)$ the ultraviolet
divergence, $n$ the space-time dimensionality, $\gamma_E$ the
Euler-Mascheroni constant, $\mu$ the 't~Hooft mass scale, and
\begin{equation}
\{I(p^2,m_l);J(p^2,m_l)\}
=\int_0^1\mathrm{d}x\,\{1;x\}\ln
\frac{m_l^2x+m_W^2(1-x)-p^2x(1-x)-i\varepsilon}{m_W^2}.
\end{equation}
The mass counterterms $\delta m_{ii^\prime}^{(\pm)}$ and endowed with very
important properties:
\begin{enumerate}
\item They are gauge independent.
Although $\Sigma_{ii^\prime}(\notp)$ contains several gauge-dependent terms,
they do not contribute to Eq.~(\ref{eq:delm}).
As explained in Ref.~\cite{Kniehl:2006rc}, such gauge-dependent terms cancel
the $(\notp-m_{i^\prime})^{-1}$ propagator in Eq.~(\ref{eq:DelM}) and
contribute to the wfr.
\item Equations~(\ref{eq:delmp}) and (\ref{eq:delmm}) automatically satisfy
the hermiticity constraint of the mass matrix, namely
\begin{equation}
\delta m_{i^\prime i}^{(-)}=\delta m_{ii^\prime}^{(+)*},\qquad
\delta m_{i^\prime i}^{(+)}=\delta m_{ii^\prime}^{(-)*}.
\end{equation}
\end{enumerate}
The gauge independence of $\delta m_{ii^\prime}$ is also easily verified by
inserting in Eq.~(\ref{eq:delm}) the expressions for
$\Sigma_{ii^\prime}^{L,R}(p^2)$ and $A_{ii^\prime}^{L,R}(p^2)$ given in
Refs.~\cite{Kniehl:2000rb,Kniehl:1990mq,Hempfling:1994ar}.
Alternatively, this can be established by means of Nielsen identities
\cite{Nielsen:1975fs}.
In fact, these identities were employed in Ref.~\cite{Espriu:2002xv} to show
that the $p^2$-dependent combination
\begin{equation}
m_im_{i^\prime}\Sigma_{ii^\prime}^L(p^2)+p^2\Sigma_{ii^\prime}^R(p^2)
+m_{i^\prime}A_{ii^\prime}^L(p^2)+m_iA_{ii^\prime}^R(p^2)
\label{eq:Nielsen}
\end{equation}
is gauge independent.
Inserting Eq.~(\ref{eq:LR}) in Eq.~(\ref{eq:Nielsen}) and evaluating the
resulting expression at $p^2=m_i^2$ and $p^2=m_{i^\prime}^2$, one immediately
observes that $\delta m_{ii^\prime}$ is gauge independent.

In the SM the functions $I(m_i^2,m_l)$, $J(m_i^2,m_l)$,
$I(m_{i^\prime}^2,m_l)$, and $J(m_{i^\prime}^2,m_l)$ are real when
$i,i^\prime\ne t$.
Thus, in such cases the $\re$ instruction is not necessary.
On the other hand, when $i=t$ ($i^\prime=t$), the first two (last two) develop
imaginary parts, and the $\re$ instruction tells us that only the real parts
of $I$ and $J$ are included in the definition of
$\delta m_{ii^\prime}^{(\pm)}$.
 
Inserting Eqs.~(\ref{eq:delmp}) and (\ref{eq:delmm}) in Eq.~(\ref{eq:DelM}),
we find
\begin{equation}
\Delta\mathcal{M}_{ii^\prime}^\mathrm{leg}
=\Delta\mathcal{M}_{ii^\prime}^\mathrm{wfr}
+\Delta\mathcal{M}_{ii^\prime}^\mathrm{res},
\end{equation}
where $\Delta\mathcal{M}_{ii^\prime}^\mathrm{wfr}$ is the wfr given in
Eq.~(30) of Ref.~\cite{Kniehl:2006rc}, and
\begin{eqnarray}
\Delta\mathcal{M}_{ii^\prime}^\mathrm{res}
&=&\frac{g^2}{32\pi^2}V_{il}V_{li^\prime}^\dagger\overline{u}_i(p)\left\{
a_+m_{i^\prime}i\im\left[\left(1+\frac{m_i^2}{2m_W^2}\right)(I-J)(m_i^2,m_l)
\right.\right.
\nonumber\\
&&{}-\left.\frac{m_l^2}{2m_W^2}(I+J)(m_i^2,m_l)\right]
+a_-m_i\left[\left(1+\frac{m_{i^\prime}^2}{2m_W^2}\right)
((I-J)(m_i^2,m_l)\right.
\nonumber\\
&&{}-\left.\left.\re(I-J)(m_{i^\prime}^2,m_l))
-\frac{m_l^2}{2m_W^2}((I+J)(m_i^2,m_l)-\re(I+J)(m_{i^\prime}^2,m_l))
\vphantom{\frac{m_i^2}{2m_W^2}}\right]\right\}
\nonumber\\
&&{}\times\frac{1}{\notp-m_{i^\prime}}
\label{eq:DelMres}
\end{eqnarray}
is a residual contribution that arises because the $I$ and $J$ functions are
evaluated at $p^2=m_i^2$ in $\Sigma_{ii^\prime}(\notp)$ and
$\delta m_{ii^\prime}^{(+)}$ [cf.\ Eqs.~(\ref{eq:DelM}) and (\ref{eq:delmp})],
at $p^2=m_{i^\prime}^2$ in $\delta m_{ii^\prime}^{(-)}$ [cf.\
Eq.~(\ref{eq:delmm})], and only their real parts are included in the
counterterms.
When $i,i^\prime\ne t$, the $I$ and $J$ functions are real in the SM and
Eq.~(\ref{eq:DelMres}) greatly simplifies:
the $a_+$ component vanishes and the $a_-$ component involves differences of
real functions evaluated at $p^2=m_i^2$ and $p^2=m_{i^\prime}^2$.

It is important to note that $\Delta\mathcal{M}_{ii^\prime}^\mathrm{res}$ is
finite and gauge independent.
Furthermore, it is non-singular in the limit $m_{i^\prime}\to m_i$, provided
that $m_i<m_W$.\footnote{%
This does not preclude the possibility of a mass-degeneracy singularity
involving two quarks with the same charges and masses $m_i,m_{i^\prime}>m_W$.
However, this hypothetical scenario is not realized in the SM with three
generations.} 
In contrast, $\Delta\mathcal{M}_{ii^\prime}^\mathrm{wfr}$ is gauge dependent
and divergent, a standard property of wfrs.
However, as explained in Refs.~\cite{Kniehl:2006bs,Kniehl:2006rc}, its
contribution to the physical $W^+\to u_i+\overline{d}_j$ amplitude does not
involve CKM matrix elements except for an overall factor $V_{ij}$, and only
depends on the masses $m_i$ and $m_j$ of the external particles, in complete
analogy with the proper vertex corrections.
As a consequence, the proof of finiteness and gauge independence of the
$W^+\to u_i+\overline{d}_j$ amplitude is reduced to that in the unmixed,
single-generation case.

For an incoming up-type quark of flavor $i^\prime$, mass $m_{i^\prime}$, and
four-momentum $p$, the external-leg correction is obtained by multiplying
$\Sigma_{ii^\prime}(\notp)-\delta m_{ii^\prime}$ by $u_{i^\prime}(p)$ on the
right and by $(\notp-m_i)^{-1}$ on the left, where $i$ denotes now the virtual
up-type quark of flavor $i$ and mass $m_i$, and
$\Sigma_{ii^\prime}(\notp)-\delta m_{ii^\prime}$ is the same amplitude
discussed before.
It is then easy to see that the residual contributions in the incoming case
are obtained by interchanging $a_+\leftrightarrow a_-$ and
$m_i\leftrightarrow m_{i^\prime}$ between the curly brackets of
Eq.~(\ref{eq:DelMres}), and multiplying the resulting expression by
$u_{i^\prime}(p)$ on the right hand and by $(\notp-m_i)^{-1}$ on the left.
Similarly, the wfr for an incoming up-type quark of flavor $i^\prime$ is
obtained by interchanging $a_+\leftrightarrow a_-$ and
$m_i\leftrightarrow m_{i^\prime}$ between the curly brackets of Eq.~(30) in
Ref.~\cite{Kniehl:2006rc} and multiplying the resulting expression by
$u_{i^\prime}(p)$ on the right.
Finally, the expressions for an outgoing down-type quark of flavor $j$ are
obtained from those of an outgoing up-type quark by substituting $i\to j$,
$i^\prime\to j^\prime$, and
$V_{il}V_{li^\prime}^\dagger\to V_{jl}^\dagger V_{lj^\prime}$, where
$j^\prime$ is the flavor of the virtual down-type quark and $l$ that of the
up-type quark in the loop.
In the case in which the external particle is a down-type quark, the $I$ and
$J$ functions are real, and the $\re$ instruction in Eqs.~(\ref{eq:delm}),
(\ref{eq:delmp}), and (\ref{eq:delmm}) is not necessary.

As discussed in Refs.~\cite{Kniehl:2006bs,Kniehl:2006rc}, diagonalization of
the complete mass matrices for both up-type and down-type quarks generates a
CKM counterterm matrix that is gauge independent, preserves unitarity in the
sense that both the bare and renormalized CKM matrices are unitary, and leads
to renormalized amplitudes that are non-singular in the limit
$m_{i^\prime}\to m_i$ for $m_i<m_W$.
A comparative analysis of the calculations of the $W$-boson hadronic widths in
various CKM renormalization schemes, including the ones proposed here and in
Refs.~\cite{Kniehl:2006bs,Kniehl:2006rc}, is presented in 
Ref.~\cite{Almasy:2008ep}.

In summary, we have presented a novel mass counterterm for CKM
renormalization that is endowed with very attractive features:
\begin{enumerate}
\item It is expressed in terms of the invariant self-energy functions, a
property that is very useful for practical applications, since such functions
are routinely evaluated in computer codes.
\item It is gauge independent, which is a crucial property to ensure the gauge
independence of the associated CKM counterterm matrix.
\item It leads to renormalized amplitudes that are non-singular in the limit
$m_{i^\prime}\to m_i$ for $m_i<m_W$.
\item It automatically satisfies the hermiticity constraints of the mass
matrix, a property that eliminates the need for special and somewhat arbitrary
adjustments of the counterterms in specific transition channels.
In fact, the counterterm presented in Eqs.~(\ref{eq:delmpm})--(\ref{eq:delmm})
can be applied as it stands to all diagonal and off-diagonal CKM amplitudes
and, in this sense, it is flavor-democratic since it does not single out
particular flavor channels.
\end{enumerate}

We thank A. Almasy for very useful discussions at the initial stage of this
study.
We are grateful to the Max Planck Institute for Physics in Munich for its
hospitality during a visit when this manuscript was prepared.
This work of was supported in part by the German Research Foundation through
the Collaborative Research Center No.\ 676 {\it Particles, Strings
and the Early Universe---the Structure of Matter and
Space-Time}.
The work of A.S. was supported in part by the National Science Foundation
Grant No.\ PHY--0245068.

\end{document}